\documentclass[sigconf,nonacm]{acmart}  

\usepackage[nolist]{acronym}        
\usepackage{algorithm}
\usepackage[noend]{algpseudocode}
\usepackage{cleveref}
\usepackage{tikz}
\usetikzlibrary{arrows.meta}
\usepackage{xspace}

\graphicspath{{figs/}}

\newcommand{\bq}{\begin{equation}}
\newcommand{\eq}{\end{equation}}
\newcommand{\bytes}{\mbox{B}}
\newcommand{\byte}{\mbox{byte}}
\newcommand{\second}{\mbox{s}}

\newcommand{\cycles}{\mbox{cy}}

\newcommand{\iter}{\mbox{it}}
\newcommand{\cyit}{\cycles/\iter}
\newcommand{\bit}{\mbox{bit}}

\newcommand{\BC}{\mbox{B/cy}}
\newcommand{\GBS}{\mbox{GB/\second}}

\newcommand{\GHZ}{\mbox{GHz}}

\newcommand{\MiB}{\mbox{MiB}}
\newcommand{\KiB}{\mbox{KiB}}

\newcommand{\likwidperfctr}{\texttt{likwid-perfctr}}
\newcommand{\likwidbench}{\texttt{likwid-bench}}
\newcommand{\likwidpin}{\texttt{likwid-pin}}

\newcommand{\PCG}{PCG\xspace}
\newcommand{\DAXPBY}{\textsc{daxpby}\xspace}
\newcommand{\DOT}{\textsc{dot}\xspace}
\newcommand{\NORM}{\textsc{norm}\xspace}
\newcommand{\STENCIL}{\textsc{stencil}\xspace}
\newcommand{\GS}{\textsc{gs}\xspace}
\newcommand{\GSfwd}{$\textsc{gs}_{\mathrm{\textsc{f}}}$\xspace}

\newcommand{\GHcomm}[1]{{\color{black}{#1}\color{black}}}
\newcommand{\JHcomm}[1]{{\color{black}{#1}\color{black}}}

\renewcommand{\Comment}[2][.5\linewidth]{%
	\leavevmode\hfill\makebox[#1][l]{->~#2}}

\begin{document}

\begin{acronym}[DVFS]
    \acro{AGU}{address-generation unit}
    \acro{AVX}{advanced vector extensions}
    \acro{CA}{cache agent}
    \acro{CL}{cache line}
    \acro{CoD}{cluster-on-die}
    \acro{DCT}{dynamic concurrency throttling}
    \acro{DIR}{directory}
    \acro{DP}{double precision}
    \acro{DP}{double precision}
    \acro{ECM}{execution-cache-memory}
    \acro{EDP}{energy-delay product}
    \acro{ES}{early snoop}
    \acro{FMA}{fused multiply-add}
    \acro{FP}{floating-point}
    \acro{HA}{home agent}
    \acro{HS}{home snoop}
    \acro{LFB}{line fill buffer}
    \acro{LLC}{last-level cache}
    \acro{LLC}{last-level cache}
    \acro{MC}{memory controller}
    \acro{MSR}{model specific register}
    \acro{NT}{non-temporal}
    \acro{NUMA}{non-uniform memory access}
    \acro{OSB}{opportunistic snoop broadcast}
    \acro{RAPL}{running average power limit}
    \acro{SIMD}{single instruction, multiple data}
    \acro{SKU}{stock keeping unit}
    \acro{SNC}{sub-NUMA clustering}
    \acro{SpMVM}{sparse-matrix-vector multiplication}
    \acro{SP}{single precision}
    \acro{SP}{single precision}
    \acro{SSE}{streaming SIMD extensions}
    \acro{TDP}{thermal design power}
    \acro{UFS}{Uncore frequency scaling}
\end{acronym}

\title[Abstracting Performance-Relevant Properties of Modern Server
Processors]{Bridging the Architecture Gap: Abstracting Performance-Relevant
Properties of Modern Server Processors}


%

\author{Johannes Hofmann}
\email{johannes.hofmann@fau.de}
\affiliation{%
    \institution{University of Erlangen-Nuremberg}
    \streetaddress{Martensstr. 3}
    \city{Erlangen}
    \country{Germany}
}

\author{Christie L. Alappat}
\email{christie.alappat@fau.de}
\affiliation{%
    \institution{University of Erlangen-Nuremberg}
    \streetaddress{Martensstr. 1}
    \city{Erlangen}
    \country{Germany}
}

\author{Georg Hager}
\email{georg.hager@fau.de}
\affiliation{%
    \institution{University of Erlangen-Nuremberg}
    \streetaddress{Martensstr. 1}
    \city{Erlangen}
    \country{Germany}
}

\author{Dietmar Fey}
\email{dietmar.fey@fau.de}
\affiliation{%
    \institution{University of Erlangen-Nuremberg}
    \streetaddress{Martensstr. 3}
    \city{Erlangen}
    \country{Germany}
}

\author{Gerhard Wellein}
\email{gerhard.wellein@fau.de}
\affiliation{%
    \institution{University of Erlangen-Nuremberg}
    \streetaddress{Martensstr. 3}
    \city{Erlangen}
    \country{Germany}
}

%
\renewcommand{\shortauthors}{Hofmann et al.}



%
\begin{abstract}

We describe a universal modeling approach for predicting single- and
multicore runtime of steady-state loops on server processors.  To this end we
strictly differentiate between application and machine models: An application
model comprises the loop code, problem sizes, and other runtime parameters,
while a machine model is an abstraction of all performance-relevant properties
of a CPU.

We introduce a generic method for determining machine models, and present
results for relevant server-processor architectures by Intel, AMD, IBM, and
Marvell/Cavium.  Considering this wide range of architectures, the set of
features required for adequate performance modeling is surprisingly small.

To validate our approach, we compare performance predictions to empirical data
for an OpenMP-parallel preconditioned CG algorithm, which includes compute-
and memory-bound kernels.  Both single- and multicore analysis shows that the
model exhibits average and maximum relative errors of 5\% and 10\%.
Deviations from the model and insights gained are discussed in detail.

\end{abstract}


%
%
%
%
%
%

\maketitle


\section{Introduction}


The architectural differences among processor models of different vendors (and even
among models of a single vendor) lead to a diverse server-processor landscape 
in the high-performance computing market.
On the other hand, several analytic
performance models, such as the Roof{}line model~\cite{hockney89,roofline:2009}
and the \ac{ECM} model~\cite{Hager:2012,Hofmann:2018},
show that many relevant performance features can be
described using a few key assumptions and a small set of numbers such as
bandwidths and peak execution rates. In
this work we introduce a structured method of establishing and describing those
assumptions and parameters that best summarize the features of a multicore
server processor. It has satisfactory predictive power in terms of
performance modeling of (sequences of) steady-state loops but is still simple
enough to be carried out with pen and paper. The overarching goal is to allow
comparisons among microarchitectures not based on benchmarks alone, which have
narrow limits of generality, but based on abstract, parameterized performance
models that can be used to attribute performance differences to one or a few
parameters or features. As a consequence, reasoning about code performance from
an architectural point of view becomes rooted in a scientific process.\medskip

\noindent\textbf{Main contributions}

\noindent
We describe an abstract workflow for predicting the runtime and
performance of sequential and parallel steady-state loops (or
sequences thereof) with regular access patterns on multicore server
CPUs. The core of the method is \GHcomm{an abstract formulation of} the \ac{ECM}
model, which is currently the only analytic model capable of
giving accurate single- and multicore estimates.

We show that a separation between the \emph{machine model}, which contains
  hardware features alone, and the \emph{application model}, which comprises
  loop code and execution parameters, is possible with some minor exceptions.

We describe a formalized way to establish a machine model for a processor
architecture and present results for Intel Sky\-lake SP \GHcomm{and, for the first time,
  for} AMD Epyc, IBM Power9,
  and Mar\-vell/Ca\-vi\-um Thun\-derX2 CPUs. The degree of data-transfer overlap
  in the memory hierarchy is identified as a key determinant for
  the single-core in-memory performance of data-bound code.
  
The feasibility of the approach is demonstrated by predicting runtime and
  performance of a preconditioned con\-ju\-gate-gra\-dient (PCG) sol\-ver and
  comparing estimates to empirical data for all investigated processors.
  \GHcomm{ECM predictions for the AMD, Cavium, and IBM CPUs
    have not been published before.}

  \medskip

\noindent\textbf{Outline}

\noindent This paper is structured as follows. In \Cref{sec:testbed} we 
detail our testbed and methodology.
\Cref{sec:approach} describes, in general terms, our modeling approach
including application model, machine model, and the modeling workflow.
\Cref{sec:construction} shows how machine models can be constructed by
analyzing data from carefully chosen microbenchmarks and gives results for the
four CPU architectures under consideration. In \Cref{sec:pcg} we validate the
model by giving runtime and performance predictions for a PCG solver and
comparing them to measurements.  Finally, \Cref{sec:related} puts our work in
context of existing research and \Cref{sec:conclusion} summarizes and
concludes the paper.


\section{Methodology and testbed}\label{sec:testbed}

\begin{table*}[!tb]
\centering
\caption{Key specification of testbed machines.}
\label{table:testbed}
    \begin{tabular}{l c c c c }
\toprule
        Microarchitecture       & Zen (EPYC)                & Skylake-SP (SKL)              & Vulcan (TX2)              & \textsc{Power}9 (PWR9)    \\
\midrule
        Chip Model              & Epyc 7451                 & Gold 6148                     & ThunderX2 CN9980          & 8335 GTX EP0S \\
        Supported core freqs    & 1.2--3.2\,\GHZ            & 1.2--3.7\,\GHZ                 & 2.2--2.5\,\GHZ            & 2.8--3.8\,\GHZ \\
        De-facto freq.          & 2.3\,\GHZ                  & 2.2\,\GHZ                      & 2.2\,\GHZ                 & {3.1\,\GHZ} \\
        Supported Uncore freqs  & 2.66\,\GHZ                 & 1.2--2.4\,\GHZ                 & 1.1\,\GHZ                 & N/A \\
        Cores/Threads           & 24/48                     & 20/40                         & 32/256                    & 22/88 \\
        SIMD extensions         & AVX2                      & AVX-512                       & NEON                      & VSX-3    \\
        L1 cache capacity       & 24$\times$32\,\KiB        & 20$\times$32\,\KiB            & 32$\times$32\,\KiB        & 22$\times$32\,\KiB \\
        L2 cache capacity       & 24$\times$512\,\KiB       & 20$\times$1\,\MiB             & 32$\times$256\,\KiB       & 11$\times$512\,\KiB \\
        L3 cache capacity       & 8$\times$8\,\MiB          & 27.5\,\MiB                    & 32\,\MiB                               & 110\,\MiB \\
        Memory Configuration    & 8 ch. DDR4-2666           & 6 ch. DDR4-2666               & 8 ch. DDR4-2400           & 8 ch. DDR4-2666 \\
        Theor. Mem. Bandwidth   & 170.6\,\GBS               & 128.0\,\GBS                   & 153.6\,\GBS               & 170.6\,\GBS \\
\midrule
        Compiler                & Intel icc 19.0 update 2   & Intel icc 19.0 update 2       & armclang 19.0             & xlc 16.1.0\\
        Optimization flags      & \texttt{-O3 -xHost}   & \texttt{-O3 -xCORE-AVX512}  & \texttt{-Ofast -mtune=native} & {\texttt{-O5 -qarch=pwr9}}\\
                                & \texttt{-mavx2 -mfma}     & \texttt{-qopt-zmm-usage=high}  &                               & \texttt{-qsimd=auto}\\
\bottomrule
\end{tabular}
\end{table*}

%
%
%


In this section we point out some relevant high-level
properties, while details will be discussed later. Note that we generally take
care to run the optimal instruction mix for all benchmark
kernels (i.e., using the most recent instruction sets available on the hardware
at hand, with appropriate unrolling in place to enable optimal instruction-level
parallelism).  Compiler peculiarities are commented on where necessary.
To minimize interference from the operating system, NUMA balancing was
disabled. Transparent huge pages were used by default.
The simultaneous multi-threading feature was ignored throughout.
\GHcomm{Measurements were carried out on repeated loop traversals so 
timer resolution was not an issue. Run-to-run variations were small
(generally below 2\%) and will thus not be reported.}

An overview of the investigated processors is provided in
Table~\ref{table:testbed}.  The AMD Epyc 7451 (EPYC) has a hierarchical design
comprising four ccNUMA nodes per socket and six cores per domain. L3 cache
segments of 8\,\MiB\ each are shared among the three cores of a core complex
(CCX)\@. The Uncore of the processor (i.e., the L3 cache, memory interface,
and other I/O circuitry) is clocked at a fixed 2.66\,\GHZ. Although the cores
support the AVX2 instruction set, 32-byte (B) wide SIMD instructions are
executed in two chunks of 16\,B by only 16-B wide hardware, so that an effective
SIMD width of 16\,B applies.

Although the Intel Xeon Skylake Gold 6148 (SKL) has a base core frequency of
2.4\,\GHZ\ and a wide range of Turbo settings, we fix the clock speed to
2.2\,\GHZ\ in all our experiments in order to avoid the automatic clock-speed
reduction when running AVX-512 code \cite{Intel:xeon-SP-spec-update}\@.  The Uncore frequency is set
to its nominal value of 2.4\,\GHZ\@. These choices are not a limitation of
generality since all procedures described in this work can be carried out for
any clock-speed setting.
SKL also features a boot-time configuration option of sub-NUMA clustering
(SNC), which splits the 20-core chip into two ccNUMA
nodes, each comprising ten cores \GHcomm{(while the full L3 is still available to
 all cores)}. This improves memory-access characteristics
and is thus a recommended operating mode for HPC in our opinion.  The
last-level cache (LLC) prefetcher was turned on for the same reason.

The Cavium/Marvell ThunderX2 CN9980 (TX2) implements the ARMv8.1
ISA with 128-\bit\ NEON SIMD extensions that support double-precision
floating-point arithmetic for a peak performance of two 16-B wide FMA
instructions per cycle and core. The 32-core chip runs at a fixed
2.2\,\GHZ\ clock speed, while the L3 cache runs at half the core speed. The
victim L3 cache is organized in 2\,\MiB\ slices but shared among all
cores of the chip.

The Power9 processor used for our investigations is part of an IBM 8336 GTX data
analytics/AI node. Being an implementation of the Power ISA v3.0, the 
core supports VSX-3 (128-\bit\ wide) SIMD instructions. A 512\,\KiB\
L2 cache is shared between each pair of cores. The victim L3 cache
is segmented, with eleven slices of 10\,\MiB\ each, and
each slice can act as a victim cache for others \cite{PWR9}. 
\medskip



The \textsc{likwid} suite~\cite{likwidweb} version 4.3.3 was
used in several contexts: \likwidpin\ for thread-core affinity, \likwidperfctr\ for
counting hardware performance events, and
\likwidbench\ for low-level loop benchmarking (with customized kernels for TX2
and PWR9)\@. Instruction latency and throughput we measured using
the \texttt{ibench} tool.\footnote{\url{https://github.com/hofm/ibench}}
Where compiled code was required, we
used the compiler versions and flags indicated in \Cref{table:testbed}\@.



\section{Modeling approach}\label{sec:approach}
\JHcomm{Just like the Roof{}line model, the ECM model is an analytic
performance model for streaming loop kernels with regular data-access patterns
and a uniform amount of work per loop iteration. Unlike Roof{}line, however,
ECM favors an analytic approach.  As a result, the model can give single- and
multicore estimates with high accuracy without relying on a large number of
measurements. Moreover, the analytic nature enables the evaluation of
different hypotheses with respect to a processor's performance behavior by
investigating which of them lead to a model that best describes empirical
performance, thereby enabling deeper insights than mea\-sure\-ment-based
approaches such as Roof{}line. See Section~\ref{sec:related} for a more
thorough comparison of the models and their predictive powers.}

\JHcomm{Two major shortcomings of the ECM model concern its loose formulation
and the resulting lack of portability: In its current form, the model mixes
general first principles and Intel-specific microarchitectural behavior into a
set of rules that make it difficult to apply it to other processors.
In the following, we untangle the original model:
First, several truly general (i.e., micro\-archi\-tec\-ture-in\-depen\-dent) first
principles and their rationales are laid out.  Next, application and machine
models that address code- and microarchitecture-specific properties are
covered (in addition, we provide general instructions on how to determine machine
models for new microarchitectures in Section~\ref{sec:construction}). Finally,
the workflow of the new model is demonstrated.}


\subsection{Model assumptions}
\label{sec:modeling-approach:model-assumptions}

%
%
The model assumes that the single-core runtime is composed of different runtime
components. These include the time required to execute instructions in the
core ($T_\mathrm{core}$) and the runtime contributions that result from
carrying out the necessary data transfers in the memory hierarchy (e.g.,
$T_\mathrm{RegL1}$ the time to transfer data between the register file and the
L1 cache, $T_\mathrm{L1L2}$ for L1-L2 transfers, and so on). Depending on the
architecture, some or all of these components may overlap. The
single-core runtime estimate is therefore derived from the runtime components
by putting them together according to the architecture's overlap capabilities.

%
%
If no shared resources are involved, single-core
performance is assumed to scale linearly with the number of active cores for the
multicore estimate.  In practice, however, at least one shared resource (the
memory interface) will be involved. The model takes conflicts on
shared resources into account by modeling contention and the resulting waiting
times in an analytical way.
In the following, some particularities of modern server processors that
simplify runtime modeling are discussed.
\begin{figure*}[!tb]
    \centering
    \includegraphics[scale=1.0]{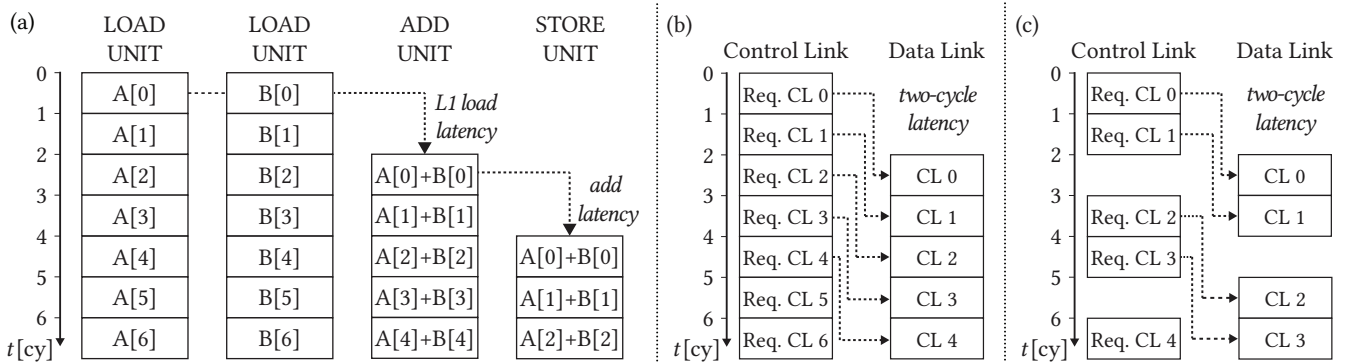}
    \caption{\label{fig:rationale-behind-assumptions}(a) Loop execution on
      a hypothetical core with load and add latencies of two cycles each.
      (b) Inter-cache data transfers for a design with
    more than two buffers to track outstanding cache-line (CL) transfers. (c) Design with only
    two buffers.}
\end{figure*}

%
%
Today's server processors typically feature superscalar, out-of-order cores
that support speculative execution \JHcomm{and implement pipe\-lined execution units.
Fig~\ref{fig:rationale-behind-assumptions}a shows the execution of
instructions corresponding to a simple vector sum
(\texttt{C[i]=A[i]+B[i]}) for a data set in the L1 cache on a hypothetical core.
The core has a two-cycle latency for add and load
instructions. When the loop begins execution, each of the two load units can
execute a load instruction. Since there is a two-cycle load latency, inputs
for the add instruction will only be available after two cycles.  However, due
to speculative execution, the core can continue to execute two load
instructions from the next loop iterations in each cycle. Once input data is
available, the core can begin executing an add instruction in each cycle.
Eventually, after another two-cycle latency (that of the add instruction), the
core can begin executing a store instruction in each cycle. Once this
latency-induced wind-up phase of four cycles is complete, instruction latency
no longer impacts runtime; instead, the runtime is determined by the
throughput of instructions. Although latencies might be higher on real
processors, the wind-up phase can be neglected even for short loops with only
hundreds of iterations. This leads to one of the key assumptions of the
ECM model: In the absence of loop-carried dependencies and data-access delays from
beyond the L1 data cache, the runtime of a single loop iteration can be
approximated by the time that is required to retire the instructions of a loop
iteration.  With loop-carried dependencies in place, the inter-iteration
critical path is a good estimate of the runtime. 
Due to speculative execution, load/store instructions are decoupled from the
arithmetic instructions of a particular loop iteration. This leads to
the further assumption that the time to retire arithmetic
instructions and the time to retire load/store instructions can
overlap.}

%
%
The next set of assumptions concerns data transfers in the memory hierarchy.
The relationship between latency and bandwidth is well understood, so most
designs typically provide a sufficient number of buffers to track outstanding
cache-line transfers to allow for the saturation of the data-transfer link
between adjacent cache levels.
\JHcomm{Fig~\ref{fig:rationale-behind-assumptions}b
shows such a design with more than two buffers to track outstanding transfers
to hide a two-cycle latency. Sometimes, however, the number of buffers is
insufficient, leading to a deterioration of bandwidth.
\Cref{fig:rationale-behind-assumptions}c shows a variant with only two
buffers: After two cycles, no more transfer-tracking buffers are available,
which prevents the initiation of new transfers. Only after a previous transfer
completes and the buffer tracking this transfer is freed can a new transfer request be
initiated. As a result, the data link is idle for one cycle, reducing the
attainable bandwidth in practice to two-thirds of the theoretical value.
On some of the investigated processors this problem can be
observed for transfers between the LLC and main memory. This
can be attributed to significant latencies caused by the increasingly complex
on-chip networks required to accommodate the growing number of cores of modern
CPUs.

The model assumes that data links can typically be fully saturated because a
sufficient amount of buffers is available and adequate prefetching (be it
hardware, software, or both) results in full utilization of these buffers. As
a result, runtime contributions of data transfers can typically be calculated
by dividing data volumes by the theoretical bandwidths of the corresponding
links; the model does, however, include an optional latency penalty to cover
edge cases such as the one shown in
Fig~\ref{fig:rationale-behind-assumptions}c.
%
%
Therefore, the  runtime contribution of data transfers between memory hierarchy
levels $i$ and $j$ is the sum of the actual data
transfer time and an optional latency penalty:
$T_{ij} = T_{ij}^\mathrm{data} + T_{ij}^\mathrm{p}$\@.}


\subsection{Application model}\label{subsec:application_model}

An application model condenses all of the code-related information
required to give runtime estimates for a particular loop.

It comprises arithmetic and load-store operations carried out during one loop
iteration as well as parameters that influence data transfers in the memory
hierarchy. Most prominently, the latter includes the data-set size(s), which
determine in which level of the memory hierarchy data resides, yet it may also
cover information about blocking size(s) and the scheduling strategy.




\subsection{Machine model}\label{subsec:machine_model}

Machine models comprise selected key information about processors.
Despite being limited to few architectural properties, the data included in
machine models is sufficient to give meaningful performance estimates.  With
respect to scope, the contents of machine models can be separated into two
parts: the execution capabilities of cores, and details about the memory
hierarchy. In the following, each of the two components is discussed in
detail.

The part concerning in-core execution capabilities deals with the cores'
properties that determine the runtime contribution of instruction execution.
As discussed in Section~\ref{sec:modeling-approach:model-assumptions}, 
throughput is a key determinant for single-core runtime, so
throughput limits (in operations per cycle) of relevant operations
are included. To address loop-carried dependencies, latencies for the
corresponding instructions must be included.
Moreover, the machine model includes
information about potential bottlenecks that limit operation throughput:
On most architectures, 
different functional units share the same execution port, which implies that
operations associated with
units served by the same port cannot 
cannot begin execution in the same cycle. Finally, most modern
core designs have one or the other peculiar shortcoming that prevents them
from fully utilizing the core's load/store units.\footnote{Most
  modern cores feature
  one store and two load units but only have two \acp{AGU}, which means that
  in each cycle only two of the three load/store units can be supplied with
  memory addresses if complex addressing modes (e.g., base plus scaled offset)
  are used.
  In addition to the two-\ac{AGU} shortcoming, the EYPC's cores have
  only two data paths between the register file and the L1 cache.}

The second part of the machine model covers information about the cache
hierarchy. This entails everything needed to calculate the volume of 
data transfers for a loop: the number of cache levels, their
\JHcomm{effective}\footnote{\JHcomm{For several reasons (imperfect cache replacement
    strategies, prefetchers preempting data that could have otherwise been reused,
    etc.) the effective capacity of a cache is lower than its
    nominal size. In practice, the heuristic of halving the theoretical
    cache size delivers good estimates for the effective size.}}
sizes, write-through vs.\
write-back policy, victim/exclusive vs.\ inclusive, etc.
For example, a victim cache typically implies
additional traffic since it receives both modified and unmodified \acp{CL}
from the overlying cache, whereas a non-victim cache only receives modified
\acp{CL}\@.
In order to get from data volumes to runtime contributions of individual
data paths, the machine model also requires data about the available bandwidth
between adjacent caches, and whether transfers take place over a
single bi-directional link or over two uni-directional links. Moreover, if an
architecture provides an inadequate number of buffers to track outstanding
transfers, the corresponding
latency penalties must be included.
Finally, the second part of the machine model contains a description of which
transfers in the memory hierarchy can occur simultaneously.\footnote{As will be
demonstrated later, we find that in practice, this rarely discussed
architectural feature turns out to be much more important for single-core
in-memory performance than other more prominent features such as SIMD width or
cache bandwidths.}


\subsection{Performance prediction workflow}
\label{subsec:application_workflow}

\begin{figure*}[tb]
    \centering
    \includegraphics[scale=1.0]{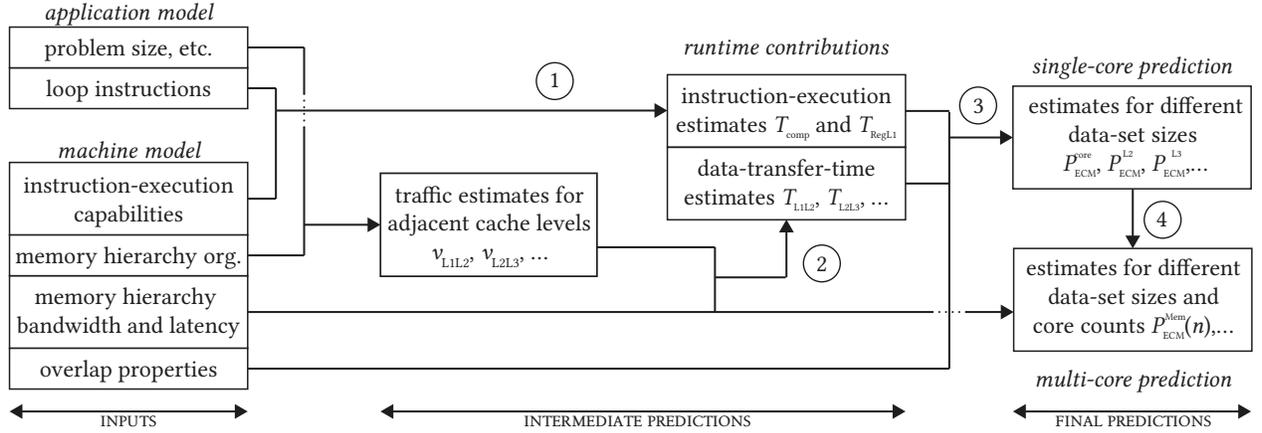}
    \caption{\label{fig:model-workflow}{Overview of the performance prediction
        workflow, including application model, machine model, and runtime
        contributions.}}
\end{figure*}
An overview of the performance-prediction workflow is provided in
\Cref{fig:model-workflow}. As indicated in the figure, the process can be
divided into four steps: First, the runtime contribution of performing
operations in the core (with all data coming from L1) is determined.
Next, the runtime contributions of
data transfers in the memory hierarchy are calculated (to this end, data transfer
volumes in the memory hierarchy need to be determined).
In a third step, the previously determined runtime contributions are
put together to arrive at a single-core runtime estimate. Finally, based on
the single-core estimate from the previous step, multicore predictions can be
given.  In the following, each of the steps is discussed in detail.

\subsubsection{Contributions of instruction execution in the core}

The fact that some architectures cannot overlap data transfers between the
register file and the L1 cache on one hand and the L1 and L2 caches on the
other makes it necessary to separate the runtime contribution of operations
into two components: $T_\mathrm{comp}$, which are cycles in which only
computational operations occur, and $T_\mathrm{RegL1}$, which are
cycles in which at least one load or store operation takes place.

To estimate $T_\mathrm{RegL1}$, first the numbers of load and store operations
($n_\mathrm{\textsc{ld}}$ and $n_\mathrm{\textsc{st}}$) are determined by
counting their occurrences in the loop body;
the numbers are then divided by the respective
throughputs, $\omega_\mathrm{\textsc{ld}}$ and
$\omega_\mathrm{\textsc{st}}$, taking additional constraints specified in the
machine model into account (e.g., a limited
throughput for the overall number of load/store operations per cycle,
$\omega_\mathrm{\textsc{ld/st}}$, caused by a limited number of \acp{AGU}). The
corresponding runtime contribution is the maximum of all components:
\begin{equation}
    \label{eq:workflow:T_RegL1}
        T_\mathrm{RegL1} =
        \max
        \left(
        \frac{n_\mathrm{\textsc{ld} }}{\mathrm{\omega}_\mathrm{\textsc{ld} }},
        \frac{n_\mathrm{\textsc{st} }}{\mathrm{\omega}_\mathrm{\textsc{st} }},
        \frac{n_\mathrm{\textsc{ld}}+n_\mathrm{\textsc{st} }}{\mathrm{\omega}_\mathrm{\textsc{ld}/\textsc{st} }}
        \right)
\end{equation}
The number of cycles in which no load/store operations are carried out is
determined in a similar way: Operation counts are found in the loop body.
Each count is then divided by the operation's throughput documented in the
machine model. As before, additional constraints have to be considered: 
For example, execution-port conflicts (cf.\ \Cref{subsec:machine_model})
can be addressed by
summing up the contributions of functional units that share the same execution
port (this is demonstrated in the equation below, where \textsc{mul} and
\textsc{div} units are assumed to be assigned to the same execution port).
The fact that cores have an upper limit to the number of instructions they can
retire per cycle can be modeled by dividing the total number of operations by a
corresponding instruction-throughput limit $\omega_\mathrm{total}$.
Finally, loop-carried dependencies are accounted for by including the
contribution of the longest cross-iteration dependency chain, $T_\mathrm{dep}$, when
determining the overall runtime by applying the maximum to all individual
contributions:
\begin{equation}
    \label{eq:workflow:T_comp}
        T_\mathrm{comp} =
        \max
        \left(
        \frac{n_\mathrm{\textsc{add} }}{\mathrm{\omega}_\mathrm{\textsc{add} }},
        \frac{n_\mathrm{\textsc{mul} }}{\mathrm{\omega}_\mathrm{\textsc{mul} }} +
        \frac{n_\mathrm{\textsc{div} }}{\mathrm{\omega}_\mathrm{\textsc{div} }},
        \ldots,
        \frac{\sum_{i} n_i}{\mathrm{\omega}_\mathrm{total}},
        T_\mathrm{dep}
        \right)
\end{equation}


\subsubsection{Contributions of data transfers in the memory hierarchy}

Before the runtime contributions of data transfers can be determined, the data
volumes transferred over the various data paths in the memory hierarchy need
to
be established. To this end, the location of the data set(s) in the memory hierarchy
is derived from the data-set size(s) specified in the application model.
Then, the load/store operations documented in the
application model are revisited: For each operation, the corresponding data
set is identified, and the transfers required to get the data from its current
location in the memory hierarchy to the L1 cache are recorded. Along with the
required transfers, the data volume is
determined (e.g., four bytes per single- or eight bytes per double-precision
floating-point number). \JHcomm{Note that full CL transfers need to be taken
  into account even when CLs are only partially read or written (e.g.,
  for strided but regular access). Moreover, the model will degenerate in case of 
  truly random access patterns as latency contributions will dominate in this
  case~\cite{cremonesi2019}\@.}
Note that this process requires keeping track of
previous data access to detect
possible data reuse.
\JHcomm{While this can be done manually for kernels with simple data-access
patterns, analysis of complex patterns is best left to cache simulators (e.g.,
pycachesim~\cite{pycachesim}). If necessary, the resulting numbers can be
validated by measuring the actual data volumes using hardware performance
events (e.g., with \textsc{papi}~\cite{PAPI} or
\textsc{likwid}~\cite{likwidweb}).}


Once the data volumes have been established, the runtime contribution $T_{ij}$
of data transfers between levels $i$ and $j$ of the memory hierarchy can be
calculated:
\begin{equation}
    \label{eq:workflow:T_ij}
        T_\mathrm{ij} =
        \mathrm{max/sum}
        \left( \frac{v_{i\rightarrow{}j}}
        {b_{i\rightarrow{}j}}, 
        \frac{v_{i\leftarrow{}j}}
        {b_{i\leftarrow{}j}} \right)
        + T_{ij}^\mathrm{p}
        = T_\mathrm{ij}^\mathrm{data} + T_{ij}^\mathrm{p}
\end{equation}
The process works by first calculating the time the data link(s) connecting
levels $i$ and $j$ are actually busy transferring data.  To calculate this
data-link busy time, $T_{ij}^\mathrm{data}$, the data volumes $v$ transferred
in each direction are divided by the bandwidth $b$ of the link over which the
data is transferred.  The two directional components $T_{i\rightarrow{}j}$ and
$T_{i\leftarrow{}j}$ are then combined according to the information provided
in the machine model. If there is a single bi-directional link over which
transfers in both directions take place, the combined data-link busy time is
the \emph{sum} of both contributions.  If there are two dedicated
uni-directional links over which the transfers can take place, the overall
data-link busy time is the \emph{maximum} of both contributions.  The
overall data-transfer time, $T_{ij}$, is given by the sum of the
previously determined data-link busy time and (if applicable) the
corresponding latency penalty specified in the machine model.


\subsubsection{Combination of runtime contributions for single-core estimate}

To arrive at a single-core runtime prediction, the previously determined
components are put together according to the overlap capabilities specified in
the machine model. To this end, first, all non-overlapping components are
added up. The result is then included in the set of overlapping components,
and the total runtime estimate is the maximum of the resulting set:
\begin{equation}
    \label{eq:runtime-estimate}
            T = \max \bigg( \overbrace{T_{\ldots}, \cdots,
    T_{\ldots},}^{\text{overlapping}}
    \overbrace{T_{\ldots} + \cdots +
    T_{\ldots}}^{\text{non-overlapping}}  \bigg)
\end{equation}
The following example will clarify the process: When
discussing the model assumptions in
Section~\ref{sec:modeling-approach:model-assumptions}, it was established that
$T_\mathrm{comp}$ and $T_\mathrm{RegL1}$ overlap on all processors. Let us
further assume that the architecture under consideration
has a multi-ported L1 cache, which enables the cache to simultaneously
communicate
with the register file and the L2 cache.  Assuming no overlap of
other transfers, the runtime
estimate for an in-memory data set on this processor would be
$T=\max(T_\mathrm{comp}, T_\mathrm{RegL1}, T_\mathrm{L1L2},
T_\mathrm{L2L3}+T_\mathrm{L3Mem}$).

The runtime estimate $T$ can be converted into a performance estimate $P$ by dividing
the amount of work $W$ carried out in one loop iteration by the runtime
estimate for the same, and multiplying the result with the core frequency:
$P = f_\mathrm{core}\cdot W/T$.

\JHcomm{For our investigations $f_\mathrm{core}$ was fixed, so converting from
runtime to performance estimates is trivial. In practice, however,
$f_\mathrm{core}$ is often set dynamically on the
authority of the operating system, the processor, or even the user.
However, $f_\mathrm{core}$ is virtually constant during
the execution of a particular steady-state loop. This is because
the metric used by the underlying mechanism (e.g., DVFS) to select
$f_\mathrm{core}$ does not change while the processor is in a steady state. For
a particular kernel, $f_\mathrm{core}$ can thus be measured via
hardware performance events.
For each kernel of a multi-loop application,
$f_\mathrm{core}$ value must be determined individually. See~\cite{Hofmann:2018} for
an investigation of the model's ability to deal with different core and Uncore
frequencies.}



\subsubsection{Multicore prediction based on single-core estimate}

Multicore estimates require as inputs the single-core runtime estimate
$T$, and the time the memory interface is busy transferring data
$T_\mathrm{Mem}^\mathrm{data}$, which is the sum of all data-link busy times
that involve the main memory (e.g., in a memory hierarchy with a
victim L3 cache, where memory sends data to L2 and receives modified \acp{CL}
from L3, $T_\mathrm{Mem}^\mathrm{data} = T_\mathrm{L2Mem}^\mathrm{data} +
T_\mathrm{L3Mem}^\mathrm{data}$).

In the absence of shared resources (e.g., if the entire data set fits into
core-private or scalable\footnote{Scalable means a parallel efficiency close
to one for all relevant degrees of parallelization (i.e., up the maximum
number of cores sharing the cache).} shared caches),
single-core performance $P$ is expected to scale linearly with the
number of active cores $n$, so the multicore estimate for $n$ active cores is
just $P(n) = nP$\@.  If shared resources, such as the main memory interface,
are involved, resource conflicts and the resulting waiting times must be
considered.  Here we employ a statistical model that is motivated by first
principles: The utilization of the memory bus $u$ is the probability of another
core encountering a busy bus. For a single core, the utilization is given by the
ratio of the time the memory interface is busy transferring data and the overall
runtime estimate: $u(1) = T_\mathrm{Mem}^\mathrm{data}/T$. If multiple cores are
active, the utilization is expressed recursively:
\begin{equation}
    \label{eq:multi-core-full}
    u(n) =
    \min \Bigg(
    1,
    \frac{nT_\mathrm{Mem}}
    { \max (T_\mathrm{comp}, \ldots, T_\mathrm{Mem} + \underbrace{u(n-1)(n-1)p_0)}_{T_\mathrm{conf}} }
    \Bigg)
\end{equation}
In the numerator, the memory-bus busy time is multiplied with the number of
active cores $n$ since multiple cores are using the memory
interface. The denominator is the expanded expression for the runtime
estimate $T$, where a conflict time has been added to the
data-transfer time involving the memory interface. This conflict time
represents the average time that a core encountering a busy memory
bus has to wait for the bus to become available to it.  The
conflict time encountered in a scenario with $n$ active cores is given by
multiplying the probability of a core hitting a busy memory bus, which
corresponds to the memory utilization of the remaining cores, $u(n-1)$, with
the time the other $n-1$ cores are using the interface. This results in
$T_\mathrm{conf}= u(n-1)(n-1)p_\mathrm{0}$, with $p_\mathrm{0}$ being an empirical
fit parameter.%
\footnote{Although $p_\mathrm{0}$ can also be modeled analytically
employing the data used to derive $T_\mathrm{Mem}$, we find that the level of
detail required to reliably estimate the parameter outweighs the benefits of
using an analytical approach.}

For performance estimates, the memory-bus utilization is multiplied with
the performance to be expected with fully saturated bandwidth:
$P(n) = u(n)P^\mathrm{Sat}$. The memory-saturation performance,
$P^\mathrm{Sat}$, corresponds to the bandwidth limitation of the
Roofline model and is determined by dividing the amount of work per loop
iteration by the memory-bus busy time, and multiplying the result with the
core frequency: $P^\mathrm{Sat} = W/T_\mathrm{Mem} \cdot f_\mathrm{core}$\@.




\section{Machine model construction}\label{sec:construction}

\subsection{Method to determine machine models}

In the ideal case, all of the data required for a machine model would be
available in vendor data sheets. In practice, however, this is rarely the case
because important information is deemed irrelevant or, more likely, intellectual
property and therefore omitted from specifications.
Moreover, the
interaction of different parts of the processor might lead to situations in
which vendor-specified numbers are not attainable (see, e.g., the discussion
on load/store throughput in Section~\ref{subsec:machine_model}).
In the following, a method is presented that allows to establish machine
models in cases where relevant information is missing, or the documented
specifications turn out to be impractical for some reason. 

\subsubsection{Instruction throughput and latency}

At fixed core clock speed $f_\mathrm{core}$, the time $t$ it takes
the core to execute a large number $n$ of independent\footnote{Independent
means that there are no data dependencies between the
different instances of the instruction.} instructions of
type $i$ is measured. 
The throughput of the
instruction is then $\tau_i=n/(t f_\mathrm{core})$. Since we will usually
use a work unit of one (high-level) loop iteration in the modeling procedure, the
instruction throughput is multiplied by the appropriate SIMD
width $w_\mathrm{\textsc{simd}}$ to get the \emph{operation throughput}
\bq
\omega_i = w_\mathrm{\textsc{simd}} \times \tau_i = w_\mathrm{\textsc{simd}}
\times n/(t f_\mathrm{core})~.
\eq


To measure latency, an artificial data-dependency chain is introduced
by making each instruction use the output of the previous instruction as its
input.  This forces each new instructions to be held at a reservation station
until the previous instruction has completed. The holding time corresponds to
the instruction's latency.

While implementing these two strategies sounds simple in theory, deriving a
suitable instruction mix from a high-level language implementation can be
difficult in practice because compiler optimizations get in the way.  We solve
this problem by side-stepping the compiler and hand-crafting the necessary
code in assembly language.  To automate the process, the \texttt{ibench}
tool
was developed, which comprises a measurement framework and a number of
assembly-code files for the most widespread instructions of AMD, IBM, ARM, and
Intel processors.


\subsubsection{Topology and data flow in the memory hierarchy}

Information about the topology of the memory hierarchy, such as the number of
caches, their sizes and properties (write-back vs.\ -through, victim vs.\
non-victim) are often well documented in vendor data sheets. Even
if this is not the case, the data is easy to obtain, for most
processors provide access to it over a well-defined interface. In case of
x86, for instance, the \texttt{cpuid} instruction can be
used to extract detailed information about the memory hierarchy, including the
capacity, associativity, number of sets, inclusiveness, cache-line size, and
more for each level in the hierarchy. Other processors offer similar
mechanisms, and the Linux \texttt{sysfs} file system provides an
architecture-independent interface to obtain the necessary data.

Information about data flow (i.e., the path data takes from a particular level
in the memory hierarchy to reach a core's L1 cache) can be derived
from the topology information. In most cases, only stores require special
attention to determine whether store-misses trigger a write-allocate for the
missed \ac{CL} or if some optimization (as the one implemented in
Marvell's ThunderX2) detects whether a full \ac{CL} is written to avoid the
write-allocate. Such details can be derived with the help of hardware
performance events, which can be used to record the data volumes exchanged
between different levels of the memory hierarchy.


\subsubsection{Bandwidth, latency, and overlap in the memory hierarchy}

In most cases, only the bandwidths of selected caches are documented by
vendors. Cache bandwidths can be determined by selecting a set of reasonable
bandwidth candidates (e.g., 16, 32, and 64\,B/cy), and examining which of the
corresponding estimates best agrees with empirical data. To the best of our
knowledge, no vendor publishes data on the overlap properties of their
processors' memory hierarchies, so this data needs to be determined in
a similar way.

The process of comparing estimates to empirical data is
iterative: Once the bandwidth and overlap properties for a particular
memory level have been established, the numbers can be used as input
for different bandwidth and overlap assumptions  in the next memory
level.
In the following the process is
demonstrated on the SKL processor for the well-known \textsc{stream}
triad~\cite{McCalpin:1995}\@.
\begin{figure*}[tb]
    \centering
    \includegraphics[scale=1.0]{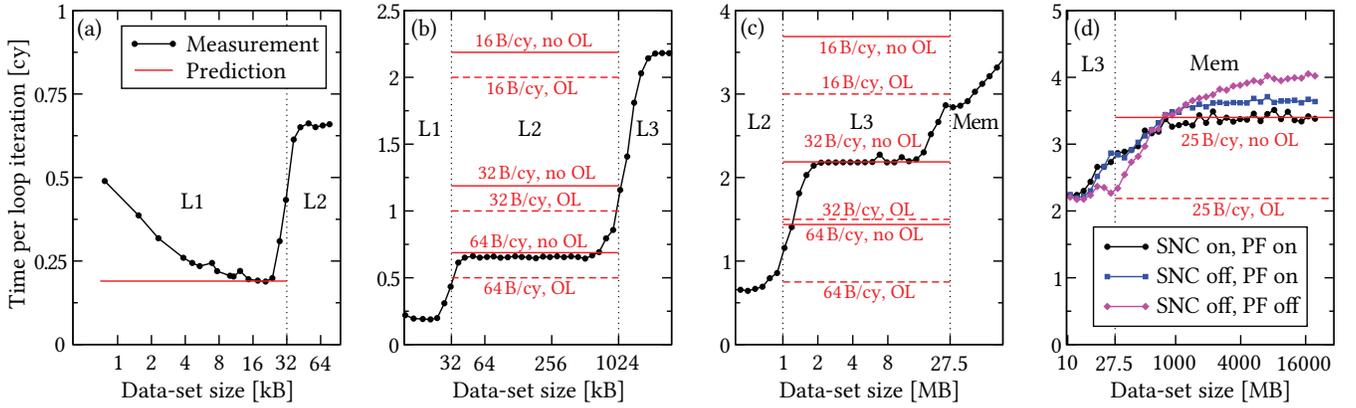}
    \caption{\label{fig:det-bw-and-ol}Comparison of model
    estimates to empirical data for the \textsc{stream} triad on SKL for data sets
    in (a) L1, (b) L2, and (c) L3 caches, and (d) main memory.}
\end{figure*}

On the SKL processor, one loop iteration of the \textsc{stream} triad
(\texttt{A[i]=B[i]+s*C[i]}) comprises two loads (one from each of the input
arrays \texttt{B} and \texttt{C}), one \ac{FMA} (to calculate the result), and
one store (to write the result to the output array \texttt{A}). Using
\texttt{ibench}, the following operation throughputs were
established:
$\omega_\mathrm{\textsc{fma}}=16/\mathrm{cy}$,
$\omega_\mathrm{\textsc{ld}}=16/\mathrm{cy}$,
$\omega_\mathrm{\textsc{st}}=8/\mathrm{cy}$, and
$\omega_\mathrm{\textsc{ld/st}}=16/\mathrm{cy}$. According to
\GHcomm{equations}~(\ref{eq:workflow:T_RegL1}), (\ref{eq:workflow:T_comp}),
and~(\ref{eq:runtime-estimate}), for a data set in the L1 cache this
leads to a single-iteration runtime estimate of
\begin{eqnarray*}
    \label{eq:application-of-runtime-estimate}
            T_\mathrm{L1} & = & \max \bigg(
            \overbrace{
                \frac{1\,\mathrm{\textsc{fma}}/\iter }{16\,\mathrm{\textsc{fma}/cy}}
            }^\text{$T_\mathrm{comp}$},
            \overbrace{
            \frac{2\,\mathrm{\textsc{ld}}/\iter }{16\,\mathrm{\textsc{ld}/cy}},
            \frac{1\,\mathrm{\textsc{st}}/\iter }{8\,\mathrm{\textsc{st}/cy}},
            \frac{3\,\mathrm{\textsc{ld/st}}/\iter }{16\,\mathrm{\textsc{ld/st}/cy}}
        }^\text{$T_\mathrm{RegL1}$}
            \bigg) \\
            & \approx & 0.19\,\cyit~.
\end{eqnarray*}
In \Cref{fig:det-bw-and-ol}a we compare this prediction to measurements.
Note that the estimate corresponds to the lower limit of runtime, which
is actually attained by the running code if the loop is long enough.

If the data set resides in the L2 cache, a total of 32\,B are transferred
between the L1 and L2 caches per iteration: 8\,B for each of the
double-precision floating-point numbers from the input arrays \texttt{B} and
\texttt{C}, 8\,B for the write-allocate to \texttt{A}, and 8\,B for evicting
the updated element of \texttt{A} to the L2 cache. Bandwidth assumptions of
16, 32, and 64\,B/cy yield estimates for $T_\mathrm{L1L2}$ of two, one, and
one-half cycle, respectively. \Cref{fig:det-bw-and-ol}b compares the estimates
to empirical data. The assumptions of no overlap and a bandwidth of 64\,B/cy
match the measurements strikingly well; incidentally, the L1-L2 cache
bandwidth as advertised by Intel is also 64\,B/cy\@.  With L1-L2 cache
bandwidth and overlap properties established, we can move on to the L3 cache.
The data exchanged between the L2 and L3 caches is 48\,B because each of the
three eight-\byte\ reads from L3 (two from the input arrays \texttt{B} and
\texttt{C}, one write-allocate from the target array \texttt{A}) triggers the
eviction of data replaced in the L2 cache to the victim L3. L2-L3 bandwidth
assumptions of 16, 32, and 64\,B/cy yield estimates for $T_\mathrm{L2L3}$ of
3, 1.5, and 0.75\,cy, respectively.  \Cref{fig:det-bw-and-ol}c compares
estimates derived from the different bandwidth and overlap assumptions to
empirical data for a data set in the L3 cache. In this case we find that that
assumptions of no overlap and a bandwidth of 32\,B/cy agree very well with the
measurement. Finally, for in-memory data sets, only different overlap
assumptions must be made, since the sustained memory bandwidth is determined
by measurement (55\,GB/s for one SNC domain, which for
$f_\mathrm{core}$=2.2\,\GHZ\ is 25\,B/cy)\@.  \Cref{fig:det-bw-and-ol}d
compares the resulting estimates to empirical data (black line) and we find
that in memory, too, no overlap of data transfers occurs.

In addition to runtime measurements obtained with SNC mode and the
LLC prefetcher (PF) enabled, \Cref{fig:det-bw-and-ol}d also shows data
where these features were disabled. This is to demonstrate that in some
settings, bandwidth and overlap are not sufficient to
describe the empirical behavior in a satisfying manner. Then, a
latency penalty must be added to data transfer times (see
\Cref{sec:modeling-approach:model-assumptions})\@.




\subsection{Results for investigated processors}

\begin{table*}[!tb]
\centering
    \caption{Machine models determined for the investigated processors.}
\label{table:machine-models}
    \begin{tabular}{l c c c c }
\toprule
        Microarchitecture       & Skylake-SP (SKL)              & Zen (EPYC)                & Vulcan (TX2)              & \textsc{Power}9 (PWR9)                        \\
    \midrule
        $\omega_\mathrm{\textsc{add}}$, $\omega_\mathrm{\textsc{mul}}$, $\omega_\mathrm{\textsc{fma}}$ [/cy]
                                & 16, 16, 16                    & 4, 4, 4                   & 4, 4, 4                   & 4, 4, 4                                       \\
        $\omega_\mathrm{\textsc{ld}}$, $\omega_\mathrm{\textsc{st}}$, $\omega_\mathrm{\textsc{ld/st}}$ [/cy]  
                                & 16, 8, 16                     & 4, 2, 4                   & 4, 2, 4                   & 4, 4, 4                                       \\
        $\lambda_\mathrm{\textsc{add}}$, $\lambda_\mathrm{\textsc{mul}}$, $\lambda_\mathrm{\textsc{fma}}$
                                & 4, 4, 4                       & 3 ,4 ,5                   & 6, 6, 6                   & 6, 6, 6                                          \\
    \midrule
        $b_{\mathrm{L1}\leftrightarrow\mathrm{L2}}$
                                & 64\,B/cy                      & 32+32\,B/cy               & 64\,B/cy                  & 64+16\,B/cy \\
        $b_{\mathrm{L2}\leftrightarrow\mathrm{L3}}$
                                & 32\,B/cy                      & 32\,B/cy                  & 32\,B/cy                  & 32\,B/cy    \\
        $b_{\mathrm{*}\leftrightarrow\mathrm{Mem}}$
                                & 25--28\,B/cy                  & 13-16\,B/cy                  & 47-56\,B/cy                  & 41-45\,B/cy    \\                                            
        Data-transfer penalties & ---                           & ---                       & ---                       & $T_\mathrm{Mem}^\mathrm{p}=0.04\,\mathrm{cy/B}$   \\
    \midrule
        Non-overlapping transfers & all                           & L2-L3, L2-Mem, L3-Mem     & all, if Mem involved                         &        L2-Mem,L3-Mem                                       \\
        Write-through/victim caches
                                & Victim L3                     & Victim L3                 &  Victim L3                & Write-through L1, Victim L3                   \\
\bottomrule
\end{tabular}
\end{table*}
	

Table~\ref{table:machine-models} shows the machine models that result from
applying the previously introduced method to the processors from the testbed.

The upper part of the table lists relevant operation throughput ($\omega$) and
instruction latency ($\lambda$) values. The center part lists bandwidths and
latency penalties (if applicable) in the memory hierarchy. Note that in cases
where two numbers are provided (e.g., 64+16\,B/cy for PWR9's L1-L2 bandwidth),
two uni-directional data paths exist between the caches. In such instances,
the first number corresponds to the bandwidth of sending data from the
underlying to the overlying cache, and second number to the bandwidth in the
opposite direction.  Note that listed memory bandwidth corresponds to
that of a single NUMA node (SNC node on SKL, Zeppelin on EPYC, full-chip on
TX2 and PWR9). Memory bandwidths are specified as ranges, since different
data-access patterns exhibit slightly varying sustained memory bandwidths.
The last part of the table contains overlap capabilities and
additional information on cache types.



\section{Case study: \PCG}\label{sec:pcg}

We use a matrix-free PCG solver to demonstrate the viability of our approach in
real-world scenarios. The solver is preconditioned using the well-known
symmetric Gauss-Seidel iteration and relies on the second-order finite-difference method
for discretization. We use it to solve the steady-state
heat equation in 2D.  The sparse matrix entries are not stored
explicitly but hard-coded into a 2D five-point stencil representation.
\GHcomm{Hence, the solver is similar to the well-known HPCG but
  shows a more interesting phenomenology: As opposed to HPCG, where all loops
  are limited by data transfers due to explicit matrix storage,
  our preconditioner is bound by in-core pipeline hazards.}
All computations and data
storage are in double precision.

Algorithm~\ref{alg:pcg} shows the entire \PCG~method. It is composed of a
matrix-free \ac{SpMVM} which we refer to as \STENCIL, a symmetric Gauss-Seidel
pre-conditioner (\GS), and three BLAS-1 routines: \DOT~product, vector \NORM,
and \DAXPBY. The code is implemented in C++ and parallelized with OpenMP\@.
The Gauss-Seidel kernels, which have loop-carried dependencies, are
parallelized using a well-known wavefront technique that preserves the
numerical behavior of the serial code~\cite{hpc4se}\@. \GHcomm{The
  preconditioner can be vectorized by, e.g., coloring methods, but this would
  alter the convergence and render the loops data bound, which is not the
  scenario we want do showcase (see above).}
\begin{algorithm}[tb]
	\caption{PCG algorithm: Solve for $x: Ax = b$}\label{alg:pcg}
	\begin{algorithmic}[1]
		\State $r = b-Ax$
        \State $r_\mathrm{norm} = \langle r,r \rangle$
		\State $p = z = Pr$
		\State $\alpha_0 = \langle r,z \rangle$
		\State $i= 0$
        \While {$(i<n_\mathrm{iter})$ \&\& $(r_\mathrm{norm}>\varepsilon^2)$}
			\State $v = Ap$ \Comment{\textsc{stencil} operation (SpMVM)}
			\State $\lambda = \frac{\alpha_0}{\langle v,p \rangle}$ \Comment{\textsc{dot}}
			\State $x = x + \lambda p$ \Comment{\textsc{daxpby}}
			\State $r = r - \lambda v$ \Comment{\textsc{daxpby}}
            \State $r_\mathrm{norm} = \langle r,r \rangle$ \Comment{\textsc{norm}}
			\State $z = Pr$ \Comment{\textsc{gs} preconditioner}
			\State $\alpha_1 = \langle r,z \rangle$ \Comment{\textsc{dot}}
			\State $p = z + \frac{\alpha_1}{\alpha_0}p$ \Comment{\textsc{daxpby}}
			\State $\alpha_0 = \alpha_1$
			\State $i=i+1$
		\EndWhile
	\end{algorithmic}
\end{algorithm}

\subsection{Application models}\label{subsec:application_model_pcg}

%
%

The total problem size ($n_i\times n_j$) was chosen to be $n_i=25000$ (inner,
leading dimension) and $n_j=2000$ (outer dimension), so that all arrays reside
in main memory.  In the following, application models for all of the PCG
components are presented.

Features important for the considered example include the 
number of loads and stores, floating-point operations, and loop structures. 
For simple streaming loops, all of these details can be derived from
high-level code.
The \DAXPBY~ kernel (\texttt{y[i]=a*x[i]+b*y[i]})  entails 
two loads, one \ac{FMA}, one multiplication, and one store.
The \DOT product (\texttt{d+=x[i]*y[i]}) and \NORM (\texttt{n+=x[i]*x[i]}) have 
two and one load(s), respectively, along with an \ac{FMA}.   These
kernels can be fully and effectively vectorized by all compilers.

For kernels with cache reuse such as \STENCIL\ and \GS, reuse-distance
analysis (best done using the layer condition~\cite{datta09}),
blocking factors, parallelization strategies, and scheduling techniques
have to be taken into account. The \STENCIL kernel is shown in
Algorithm~\ref{alg:stencil}, with $w_{\ast}$ representing different weights
obtained from the matrix $A$.
\begin{algorithm}[tb]
    \caption{High-level representation of \STENCIL}\label{alg:stencil}
\begin{algorithmic}[1]
	\For {$j=1:n_j-1$}
		\For {$i=1:n_i-1$}
            \State $v_{j,i}=w_c p_{j,i} + w_y (p_{j-1,i}+p_{j+1,i})+ w_x (p_{j,i-1} + p_{j,i+1})$
		\EndFor
	\EndFor
\end{algorithmic}
\end{algorithm}
The kernel requires two \acp{FMA}, two additions, one multiplication, one
store, and five load operations.  SIMD vectorization is straightforward, but
in contrast to the BLAS kernels, different loads can hit different memory
hierarchy levels depending on the reuse distance. For the considered inner
dimension of $n_i=25000$ and outer ($j$) loop parallelization employed in our
code, the layer condition would require $4 n_i$ elements per thread to
fit in a cache.  The lowest (i.e., outermost) cache that satisfies this
criterion will only have a miss for one of the four elements on the right-hand
side, while the cache levels above it will have three.
Storing to $v$ implies a write-allocate through the whole memory hierarchy on
all processors, and, at some point, the writing back of the newly-computed data to
memory.

The \GS kernel is a symmetric operator comprising a forward and a backward
sweep.  The forward sweep (\GSfwd) is shown in Algorithm~\ref{alg:GS_fwd}, and requires two \acp{FMA},
one multiplication, one store, and three load operations.
\begin{algorithm}[tb]
    \caption{High-level representation of \textsc{gs} forward sweep}\label{alg:GS_fwd}
    \begin{algorithmic}[1]
        \For {$j=1:n_j-1$}
        \For {$i=1:n_i-1$}
        \State $z_{j,i}= w_c (r_{j,i} + w_y z_{j-1,i} + w_x z_{j,i-1})$
        \EndFor
        \EndFor
    \end{algorithmic}
\end{algorithm}
The kernel is similar to \STENCIL, but it reads from $z_{j,i-1}$ and writes to
$z_{j,i}$, causing a loop-carried dependency.  A wavefront technique can be used
to parallelize the kernel~\cite{hpc4se}, and the corresponding layer-condition criterion
requires $3 n_i$ elements to fit in a cache. The outermost cache that satisfies
this condition will have only two load misses on the right-hand side,
while the others would have three. The \GS backward sweep (not shown here for brevity)
is similar, but loops are traversed in reverse direction
and $w_c r_{j,i}$ in \GSfwd is replaced with $z_{j,i}$. The analysis of the
kernel follows the same approach, but there is one less load miss.

\begin{figure}[tb]
	\centering
	\scalebox{0.8}{
	\begin{tikzpicture}
	\begin{scope}[every node/.style={thick}]
        \node[draw] (A) at (0,0.5) {{LOAD} $r_{j,i}$};
	\node[draw] (B) at (0,2) {$w_y$};
        \node[draw] (C) at (0,4) {{LOAD} $z_{j-1,i}$};
        \node[draw,circle] (D) at (2, 1.5) {{FMA}};
        \node[draw,circle] (E) at (4, 1) {{FMA}};
	\node[draw] (F) at (2.5, 2.5) {$w_x$};
        \node[draw] (G) at (2.5, 4) {$z_{j,i-1}$};
        \node[draw,circle] (H) at (5.5, 2.5) {{MUL}};
	\node[draw] (I) at (4.5, 4) {$w_c$};
        \node[draw] (J) at (7.5, 2.5) {{STORE} $z_{j,i}$};
	\node (K) at (6.25, 2.5) {};
	\end{scope}
	
	\begin{scope}[>={Stealth[black]},
	every node/.style={fill=white,circle},
	every edge/.style={draw=black,very thick}]
	\path [->] (A) edge  (D);
	\path [->] (B) edge  (D);
	\path [->] (C) edge  (D);
	\path [->] (D) edge  (E);
	\path [->] (F) edge  (E);
	\path [->] (G) edge[red]  (E);
	\path [->] (E) edge[red]  (H);
	\path [->] (I) edge  (H);
	\path [->] (H) edge[red]  (J);
	\path [->] (K) edge[bend right=60, red]  (G);
	\end{scope}
	\end{tikzpicture}
}
    \caption{\label{fig:gs_dependency} Dependency chain of the \textsc{gs}
          forward kernel when using the Intel compiler. Critical path shown
          in red.}
\end{figure}
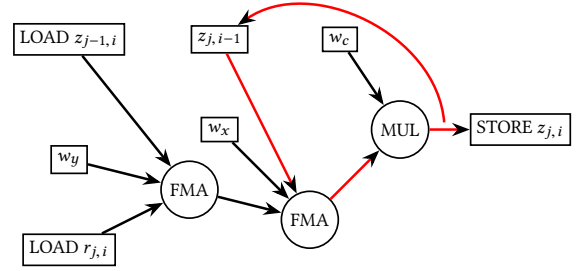

Both \GS loops have loop-carried dependencies, preventing
SIMD vec\-torization. As a result, a critical path analysis is required.  In \GSfwd
the element $z_{j,i}$ written in a particular iteration
is read in the next as $z_{j,i-1}$. The
actual delay caused by this dependency can vary depending on the code
generated by the compiler.
\Cref{fig:gs_dependency} shows 
the result when using the Intel compiler and the critical path of the generated
instruction mix includes
one \ac{FMA} and one multiplication.
The ARM clang compiler produces code that does not keep
$z_{j,i}$ in a register across loop iterations, leading to an extra
delay caused by storing and loading the element. Due to its particular
unrolling strategy, IBM's \textsc{xlc} compiler
generates a combination of the two previous variants.




\subsection{Runtime predictions}

In the following, the proposed model is validated by comparing the model
estimates to empirical performance for the \DAXPBY and \GSfwd kernels, as well as the
full PCG algorithm.  Note that estimates correspond to the runtime of a single high-level
loop iteration.

\subsubsection{Single-core}


On the SKL processor, retiring the \textsc{daxpby} kernel's multiplication and
\ac{FMA} operations takes $T_\mathrm{comp}\approx 0.0625\,\cycles$. The
one store and two load operations take $T_\mathrm{RegL1} \approx
0.1875\,\cycles$. Per-iteration data-transfer volumes
are 24\,\bytes\ between the L1 and L2 caches (one load each from \texttt{x} and
\texttt{y}, one write to \texttt{y}), 32\,\bytes\ between the L2 and L3 caches
(one load each from \texttt{x} and \texttt{y}, and two corresponding evicts
since the L3 is a victim cache), and 24\,\bytes\ between L3 and main memory
(see L1-L2 transfers)\@.
Using the bandwidths documented in the machine model, this results in
contributions of $T_\mathrm{L1L2}=0.375\,\cycles$, and
$T_\mathrm{L2L3}=1\,\cycles$. For the measured memory bandwidth of
60\,\GBS, which for $f_\mathrm{core}=2.2\,\GHZ$ corresponds to a
bandwidth of 27.3\,\BC, $T_\mathrm{L3Mem}$ is 0.88\,\cycles.  Since all data
transfers are non-overlapping, the runtime estimates are
$T_\mathrm{L1}=0.1875\,\cycles$, $T_\mathrm{L2}=0.5625\,\cycles$,
$T_\mathrm{L3}=1.5625\,\cycles$, and $T_\mathrm{Mem}=2.4425\,\cycles$.

Intermediate and final single-core estimates for \textsc{daxpy} on SKL, and
all other processors, are given in Table~\ref{tab:single-core-daxpy}. Cases
where data volumes change in the victim L3 cache (depending on whether the
input data resides in the L3 or main memory) are indicated by listing two
numbers in the table, the former corresponding to the data-transfer time
estimate for data in the L3, the latter for data in memory.

\begin{figure*}[tb]
    \centering
    \includegraphics[scale=1.0]{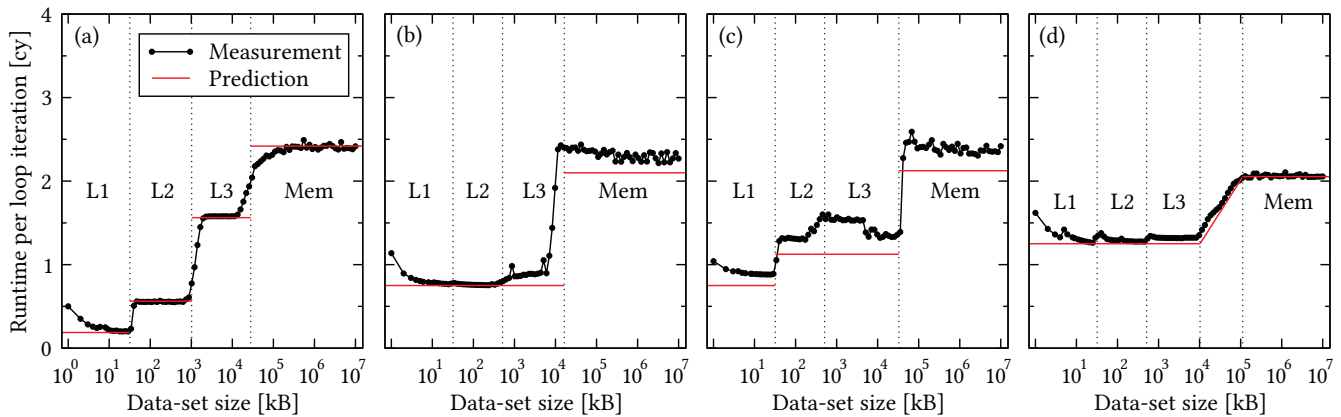}
    \caption{\label{fig:single-core-daxpy-validation}Comparison of model
    estimates to empirical data for \textsc{daxpy} on (a) SKL, (b) EPYC, (c)
    TX2, and (d) PWR9.}
\end{figure*}

These single-core estimates are compared to empirical data in
\Cref{fig:single-core-daxpy-validation}. The data indicates that the model
manages to describe empirical performance on all investigated
processors with high accuracy.

\begin{table}[!b]
\centering
    \caption{Single-core estimates for \textsc{daxpy} on all investigated
    processors.}
    \label{tab:single-core-daxpy}
    \begin{tabular}{l c c c c }
\toprule
        CPU                     & SKL              & EPYC                   & TX2               & PWR9    \\
\midrule
        $T_\mathrm{comp}$ [cy/it]      & 0.0625    & 0.25                   & 0.25          & 0.25\\
        $T_\mathrm{RegL1}$[cy/it]      & 0.1875    & 0.75                   & 0.75          & 0.75\\
        $T_\mathrm{L1L2}$ [cy/it]      & 0.375     & 0.5                    & 0.375          & 0.5\\
        $T_\mathrm{L2L3}$ [cy/it]      & 1         & 0.75 | 0.25            & 1 | 0.5       & 1 | 0.5\\
        $T_\mathrm{L2Mem}$[cy/it]      & ---       & 1.23                   & 0.29          & 0.36\\
        $T_\mathrm{L3Mem}$[cy/it]      & 0.88      & 0.62                   & 0.14          & 0.18\\
\midrule
        $T_\mathrm{L1}$   [cy/it]      & 0.1875    & 0.75                   & 0.75          & 1.25 \\
        $T_\mathrm{L2}$   [cy/it]      & 0.5625    & 0.75                   & 1.125          & 1.25 \\
        $T_\mathrm{L3}$   [cy/it]      & 1.5625    & 0.75                   & 1.125          & 1.25 \\
        $T_\mathrm{Mem}$  [cy/it]      & 2.4425    & 2.1                    & 2.06          & 2.1 \\
\bottomrule
\end{tabular}
\end{table}

\subsubsection{Multicore}
The \DAXPBY and \GSfwd kernels were selected to investigate the model's
capability to accurately describe multicore performance.
Being a data-bound streaming kernel, \DAXPBY proves particularly suitable to
investigate the memory subsystem of the investigated processors and their
scaling behavior.  \GSfwd, on the other hand, is core bound for all
architectures when executed on a single core. However, when increasing the number
of cores, NUMA properties turn out to have a significant impact on performance.


\begin{figure*}[tbp]
    \centering
    \includegraphics[scale=1.0]{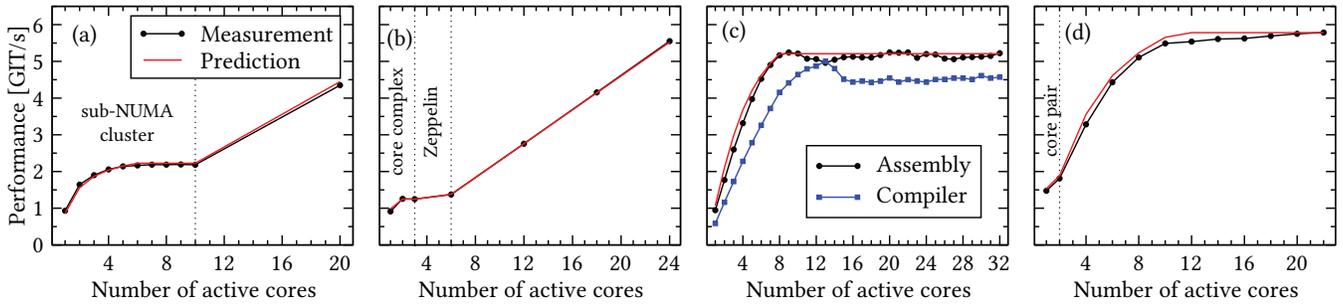}
    \caption{\label{fig:multi-core-daxpy}Comparison of 
      performance models to empirical data for intra-socket scaling of
      \DAXPBY on (a) SKL, (b) EPYC, (c) TX2, and (d) PWR9\@. Performance
      is given in $\boldsymbol{10^9}$ iterations per second (GIT/s).}
\end{figure*}
\Cref{fig:multi-core-daxpy} shows the multicore scaling of \DAXPBY on
all architectures up to a full socket using ``close'' thread affinity (i.e.,
filling cores consecutively through ccNUMA domains)\@.
For SKL we observe the typical saturation behavior (at
$\approx$ 2.2\,GIT/s = 53\,\GBS) of bandwidth-bound code within a single
SNC domain.  Using the second SNC domain doubles the bandwidth and hence
performance by a factor of two as predicted by the model. The scaling behavior of
EPYC exposes its main hardware features: Within a single CCX (three cores) the shared L3
bandwidth does not scale across the cores and hits a maximum
of 32\,\BC\@. 
The best bandwidth attained on a single CCX
is 30\,\GBS\ compared to 33\,\GBS\ for the entire ccNUMA domain (a ``Zeppelin'' die);
we speculate that this is a faint echo of non-scalable L3 cache.
Scaling across the Zeppelin dies is linear as expected.
\JHcomm{On the TX2, we initially observed a significant deviation:
  The compiler-generated code (blue line)
  fell short of the model by as much as 40\% for a single core and 10\% after
  saturation. The prompted investigation revealed that ARM's compiler did not generate
  prefetch instructions, which prove imperative for best
  performance of data-bound loops. Manually adding prefetching
  instructions to the compiler-generated code brought model and
  measurement together (black line).  This demonstrates how the model can be
  used to identify bottlenecks or other shortcomings that limit performance (in
  this case, the compiler).  Note that the optimization is not part of our PCG
  code; we use the compiler versions for all further comparisons.}
On PWR9, the scaling within a core pair is similar to that observed within a CCX
of EPYC.  This is due to the shared and non-scalable L2 and L3 cache segments
per core.  The multicore model accommodates this behavior by keeping the L2 and
L3 data-transfer rates constant for the two cores sharing the resources.  
Scaling across core pairs (i.e., running with 2, 4, 6, etc. cores)
is only limited by bandwidth saturation as can be
observed by the measurements and respective model prediction.

\begin{figure}[tbp]
    \centering
    \includegraphics[scale=1.0]{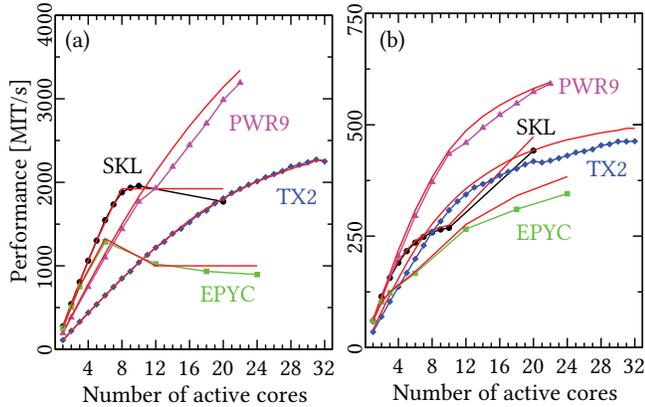}
    \caption{\label{fig:multi-core-gsfwd-and-composite}Comparison of 
    estimates to empirical data for (a) GS forward kernel and (b) the full PCG
    algorithm.}
\end{figure}

The \GSfwd kernel is latency bound due to the loop-carried dependency
discussed in \Cref{subsec:application_model_pcg}.  There are two peculiarities
that make predictions of the parallel \GSfwd kernel challenging: First, the
wavefront parallelization requires a barrier synchronization after each inner
loop traversal. For the chosen problem size, the corresponding OpenMP-barrier was found
to cause non-negligible overhead.  We addressed this by benchmarking the
OpenMP barrier for all relevant compiler-hardware combinations and included
the barrier time as additional overhead. Secondly, although parallel
first-touch page placement works fine for all other loops, the
parallel-wavefront algorithm accesses data in parallel across the inner
dimension. Since data placement is done with static Open\-MP scheduling across
the outer dimension, this leads to all threads accessing the same ccNUMA
domain most of the time during the \textsc{gs} sweeps.  It turns out that this
effect can be incorporated into the model as well. To this end, the sustained
memory bandwidth is measured across all ccNUMA domains with data residing in
only one domain. This data can then be used as a bandwidth limit when using
multiple ccNUMA domains
on SKL and EPYC\@.
\Cref{fig:multi-core-gsfwd-and-composite}a compares performance estimtes
to measurements for \GSfwd across the cores of a socket on
all architectures. The deviation from the model is generally smaller than 10\%
when using multiple NUMA domains, and below 5\% when looking at a single
ccNUMA domain. The results indicate that the model with enhancements
described above (barrier overhead, ccNUMA contention) delivers a good
qualitative and quantitative description of the performance behavior.

%

\subsubsection{Composition}

With estimates for individual kernels in place we can now
present multicore-scaling data for the full PCG algorithm.  Composing
the model from single-loop predictions is simple due to the time-based
formulation of the ECM model~\cite{ECM_composition}\@.  In case of PCG
we have three invocations of \DAXPBY, two of \DOT, 
one \textsc{gs} forward- and backward-sweep each, as well as one of \STENCIL.
\Cref{fig:multi-core-gsfwd-and-composite}b shows the comparison of the
model with measurements for all four architectures. Again, the
general model error is below 10\%, and less than
5\% when looking at single ccNUMA domains. The slightly larger deviation 
beyond 12 cores on TX2 can be attributed to the fact that we use
compiler-generated code instead of hand-crafted assembly for the
CG solver on this machine. The lack of prefetching causes
a 10-15\% performance breakdown of data-bound loops beyond
the saturation point (see~\Cref{fig:multi-core-daxpy}c), which
we ignore in the model. On EPYC and SKL we observe very low
performance for OpenMP reductions across ccNUMA domains
(much larger than the considered OpenMP barrier) with the Intel
compiler, causing the slight deviation beyond one domain.




\section{Related work}\label{sec:related}

There are two capable analytic (in the sense of ``first principles'')
performance models for steady-state loop code on multicore CPUs: the
Roof{}line model~\cite{hockney89,roofline:2009} and the ECM
model~\cite{Hager:2012,sthw15,Hofmann:2018}\@. Both have been subject
to intense study, refinements, and validation, and their areas of
applicability are well understood. \GHcomm{However, while there is ample data
available for Roof{}line on a wide variety of
architectures~\cite{Ofenbeck:2014,Ilic:2014}, one drawback of previous
applications of the ECM model \cite{sthw15,Gmeiner:2015,Hofmann:2016,Wittmann:2016,ECM_composition,Wichmann:2018,cremonesi2019} is that they were mostly restricted to Intel
processors. We provide the first thorough cross-architecture study of
the model.}

The Roof{}line model has the attractive property that it can be
easily separated into a machine part (memory and cache bandwidths,
peak performance) and an application part (computational intensity)\@.
There is no previous work that has done the same with the ECM model.
A comparison between Roof{}line and ECM for several stencil
algorithms can be found in~\cite{sthw15}.
A drawback of the Roof{}line model is that it requires a large
amount of phenomenological input such as measured bandwidths for
all core counts and all memory hierarchy levels~\cite{Ilic:2014}, while
the ECM model only needs the saturated memory bandwidth and
the machine model (i.e., overlap assumptions). 

Advanced curve-fitting and machine-learning techniques combined with
hardware performance monitoring data have been used in the past to
model the performance of code~\cite{Alam:2007,Peraza:2013}. Although
these approaches are useful in practical settings, e.g., for
predicting program runtimes with a goal of optimized resource
scheduling, the deepest insights are gained through first-principles
models such as Roof{}line or ECM.


\section{Conclusion}\label{sec:conclusion}


We have shown  that it is possible to set up a
well-defined workflow for modeling the serial and parallel runtime of
steady-state (sequences of) loops with regular data access patterns
using the analytic ECM performance model. One can, with minor
exceptions, cleanly separate machine properties from application
properties. Four multicore server processors were investigated, and we could
demonstrate that despite their obvious differences the main properties
needed to set up a useful machine model can be summarized in a few
parameters. The performance, including scalability across cores and
ccNUMA domains, of an OpenMP-parallel preconditioned CG solver with
wavefront-parallel Gauss-Seidel sweeps could be described with a
modeling error of 5\% or less in most cases.

We found the overlapping property of
transfers across data paths in the cache hierarchy to be the
pivotal architectural feature governing single-core performance
for data-bound loops.  A design with
very strong in-core performance (e.g., via wide SIMD execution) but a
non-overlapping memory hierarchy may well be inferior to a weak core
with strong overlap, as our comparison of Skylake SP and AMD Epyc
shows. The architecture with the lowest in-core computational performance, Power9,
came out first in serial and parallel memory-bound performance. The
Cavium ThunderX2 processor can compensate its rather low in-core
performance with good memory bandwidth and a large core count.

\GHcomm{All modeling procedures carried out in this
  paper were done by hand. Some components, e.g., the construction
  of a runtime prediction from code and a (given) machine model,
  can be supported by tools~\cite{Hammer:2017}; others,
  such as the derivation of overlapping properties, would
  be very hard to automate. However,
  the purpose of performance modeling is not just prediction
  but also insight, and manual analysis sharpens the view on
  the relevant details.}


\begin{acks}
    We thank Thomas Gruber for helping to port the \textsc{likwid} tool suite to IBM's
    \textsc{Power}9 architecture.

    We also thank the Center for Information Services and High Performance
    Computing (ZIH) at TU Dresden for providing access to their
    \textsc{Power9} cluster.
\end{acks}

\bibliographystyle{ACM-Reference-Format}
\bibliography{pub}


\end{document}